\documentclass[12pt]{article}
\usepackage{amssymb,amsmath,epsfig,graphicx}
\begin{document}
\title{\bf Analysis of Charged Compact Stars in Modified Gravity}
\author{M. Farasat Shamir
\thanks{farasat.shamir@nu.edu.pk} and Saeeda Zia \thanks{saeeda.zia@nu.edu.pk} \\\\
Department of Sciences and Humanities, \\National University of
Computer and Emerging Sciences,\\ Lahore Campus, Pakistan.}
\date{}
\maketitle

\date{}

\maketitle
\begin{abstract}
 Current study highlights the physical characteristics of charged anisotropic compact stars by exploring some exact solutions of modified field equations in $f(R,G)$ gravity. A comprehensive analysis is performed from the obtained solutions regarding stability, energy conditions, regularity, sound velocity and compactness. These solutions can be referred to model the compact celestial entities. In particular, a compact star named, $Her X-1$ has been modeled which indicates that current solution fits and is in conformity to the observational data as well. A useful and interesting fact from this model arises that relative difference between two forces of anisotropic pressure and electromagnetic force may occur inside the aforementioned compact star. This is another mechanism which is essential for stability of the compact object and prevent stellar object to annihilate.
\end{abstract}

{\bf Keywords:}  $f(R,G)$ Gravity, Spherically Symmetric Spacetime, Compact Stars.\\
{\bf PACS:} 04.50.Kd, 98.80.-k.

\section{Introduction}

Despite of the great success of theory of  General relativity (GR) in the previous century, several useful modifications are being proposed by researchers. For example $f(R), f(G), f(R,T), f(R,G)$ and many other as well. Here, $R, G$ and $T$ are representing the Ricci scalar, Gauss Bonnet term and trace of the energy momentum tensor respectively. The Gauss Bonnet term is defined as follows:\\
\begin{equation}
G\equiv R^2-4R_{\alpha\beta}R^{\alpha\beta}+R_{\alpha\beta\theta\phi}R^{\alpha\beta\theta\phi},
\end{equation}
where $R_{\alpha\beta}$ and $R_{\alpha\beta \theta\phi}$ symbolize Ricci and Riemann tensors respectively.
The important phenomenon which is being addressed by these theories is expansion of the universe \cite{1}. It is a known fact that GR alone in its regular arrangement can not explain the accelerated expansion without involving extra expressions in the gravitational Lagrangian or exotic matter \cite{1a},\cite{1b}. In $1970$, a simplest idea of modification was given by Buchdal \cite{2} with the replacement of $R$ by $f(R)$ in the Einstein field equations. Now lot of information is available in literature regarding modified theories of gravity \cite{3}-\cite{10}. A well known theory namely, Gauss Bonnet gravity has been discussed many a times in the recent past \cite{11},\cite{12}. In these modified theories, the gravitational lagrangian comprises of a function $f(R,G)$.  This is a fact that involvement of Gauss Bonnet term can resolve short comings of the $f(R)$ theory to the augmented expansion of the universe \cite{10a}-\cite{10c}. The $f(R,G)$ gravity is widely studied and it can reproduce any type of cosmological solutions.

Observations of compact matters like pulsars, black holes and different neutron stars have moved the research away from just finding mathematical expeditions into the useful modeling physical stuffs based on highly precise observational data \cite{13}. Physical properties of three different compact stars in the framework of $f(R,G)$ gravity have been discussed recently and it was concluded that all the three stars under consideration are stable and energy conditions are also satisfied \cite{14}. For modeling static objects, assumption of spherical symmetry is quite natural while there is more freedom in the selection of matter content. For modeling static objects, assumption of spherical symmetry is quite natural while there is more freedom in the selection of matter content. In past, researchers have focused their work on perfect fluids. However, pressure anisotropic fluids and fluids with viscosity have also been considered and it has been shown that anisotropy can effect the stability of the configuration as compared to the local isotropic case. Moreover, various type of physical phenomena and some density ranges increase the local anisotropy \cite{2000}. The effect of local anisotropy in relativistic spheres using equation of state has been studied with some useful results \cite{1000}. Thus it seems appropriate to consider pressure anisotropy in particular with modified gravity models. Advanced cosmology has motivated the researchers the inclusion of dark and phantom energy in the study of stellar objects \cite{15}. In recent years, researchers have found out some exact solutions for Einstein field equations using embedding class one \cite{15a}. Karmarkar developed a Constraint which serves as a necessary condition for embedding a $4$-dimensional spherically symmetric space time into a $5$-dimensional flat space time \cite{15b}. A further obligation is also required to make sure that the Karmarkar condition is sufficient. The geometric derivation of this condition gives a connection between both gravitational potentials. This is helpful for obtaining a detailed report on the gravitational behavior of the concerned model as one requires  specific metric functions and for the other one Karmarkar condition is used. It is important to mention here that Karmarkar condition along with the pressure isotropy suggests that the Schwarzchild interior solution provides bounded matter formation with disappearing pressure anisotropy.

Study of different stars and particle physics inside their dense cores compelled the researchers for more genuine solutions of field equations. Involving pressure anisotropy and permission of charge inside the stellar objects has provided an interesting way to observe the physical properties of compact stars \cite{15b1}-\cite{15b4}. In present work, we model a charged spherically symmetric object by inserting spherically symmetric static space time having Schwarzchild coordinates into a $5$-dimensional flat space time. The pressure is assumed to be anisotropic within the distribution. The organization of this paper is as follows: In section $2$, we describe the modified field equations with charged anisotropic matter distribution. In section $3$, we find gravitational potentials by embedding class condition and a generalized model is created for the charged anisotropic solution. Section $4$ is devoted for physical analysis of the solution and finally last section concludes the paper.
\section{Field equations and charged anisotropic matter distribution in $f(R,G)$ gravity}

In Schwarzschild coordinates $(t,r,\theta,\phi)$, the interior of a spherically symmetric and static matter distribution is as follows:
\begin{eqnarray}\label{1}
ds^{2}=e^{a(r)}dt^{2}-e^{b(r)}dr^{2}-r^{2}{(d\theta^{2}+\sin^{2}\theta d\phi^{2})},
\end{eqnarray}
where we have to find the potentials $a$ and $b$ depending upon radius $r$.
In this work, we study the charged compact stars in context of $f(R,G)$ gravity. The field equations concerning the space time geometry to the matter content are
\begin{eqnarray}\label{2}
R_{\mu\nu}-\frac{1}{2}g_{\mu\nu}R&=&\kappa( T_{\mu\nu}+E_{\mu\nu})+\nabla_\mu\nabla_\nu f_R-g_{\mu\nu}\Box f_R+2R\nabla_\mu\nabla_\nu f_G-2g_{\mu\nu}R\Box f_G\nonumber
\\&& -4R^\alpha_\mu\nabla_\alpha\nabla_\nu f_G
-4R^\alpha_\nu\nabla_\alpha\nabla_\mu f_G+4R_{\mu\nu}\Box f_G+4g_{\mu\nu}R^{\theta\phi}\nabla_\theta\nabla_\phi f_G\nonumber \\&&+4R_{\mu\theta\phi\nu}\nabla^\theta\nabla^\phi f_G-\frac{1}{2}g_{\mu\nu}V+(1-f_R)G_{\mu\nu},
\end{eqnarray}
where $\kappa$ is the Einstein coupling constant and partial derivatives with respect to $R$ and $G$ are denoted by $f_R$ and $f_G$ respectively,
\begin{eqnarray}
V\equiv f_RR+f_GG-f(R,G),
\end{eqnarray}
and $T_{\mu\nu}$ describes the usual energy-momentum tensor. We work here with the assumption that radial and tangential pressures within the interior matter distribution are not equal which implies that matter within the object is locally anisotropic.
The energy-momentum tensor and electromagnetic field in case of anisotropic fluid is given by
\begin{equation}
T_{\alpha\beta}=(\rho+p_t)u_\alpha u_\beta-p_tg_{\alpha\beta}+(p_r-p_t)v_\alpha v_\beta,
\end{equation}
\begin{eqnarray}
E_{\alpha\beta}=\frac{g_{\alpha\alpha}}{4}(-F^{\alpha m}F_{\beta m}+\frac{1}{4}\delta^\alpha_\beta F^{mn}F_{mn}),
\end{eqnarray}
where $u_\alpha=e^{a/2} \delta_\alpha^0$, $v_\alpha=e^{b/2}\delta_\alpha^1$ are four velocities. Radial pressure and tangential pressures are $p_r$ and $p_t$ respectively, and the energy density is denoted by $\rho$. The components of $T_{\alpha\beta}$ and $E_{\alpha\beta}$ are found out respectively as:
\begin{eqnarray}\label{3}
T_{00}=\rho e^a, T_{11}=p_r e^b, T_{22}=r^2p_t,
\end{eqnarray}
\begin{eqnarray}\label{4}
E_{00}=\frac{Q^2e^{-b}}{8\pi r^4}, E_{11}=\frac{Q^2e^{-a}}{8\pi r^4}, E_{22}=\frac{r^2Q^2e^{-(a+b)}}{8\pi r^4},
\end{eqnarray}
As we are using spherical symmetry, the zero-current component is only a function of $r$. For the electromagnetic field tensor, the non zero components are $F^{01}$ and $F^{10}$, related by $F^{01}=F^{10}$, which shows the radial component of the electric field. The total charge confined in the sphere of radius $r$, denoted by $Q(r)$, is defined using the relativistic Gauss law as \cite{15a}
\begin{eqnarray}
Q(r)=4\int^\pi_0\sigma r^2 e^{\frac{b}{2}}dr.
\end{eqnarray}
  Using equations (\ref{3}) and (\ref{4}) in field equations (\ref{2}) and after some simplifications we get,
\begin{eqnarray}\label{5}
\rho+\frac{Q^2}{8\pi r^4}e^{-(a+b)}&=&-e^{-b} f''_{1R}-\frac{4e^{-2b}}{r^2}(a'r-b'r-e^b+4)f''_{2G}-e^{-b}(\frac{b'}{2}+\frac{2}{r})f'_{1R}+\nonumber\\
&&e^{-2b}(a''a'+ a''b'-a''-a'^2+a'b'+\frac{a'^3}{2}-\frac{4a'}{r}-\frac{4b'e^b}{r^2}+\nonumber\\
&&\frac{8a''}{r}+\frac{6a'^2}{r}-\frac{13b'}{r^2}-\frac{2b'^2}{r}-\frac{17a'}{r^2}-\frac{18e^b}{r^3}+\frac{18}{r^3})f'_{2G}+\frac{e^{-b}}{r^2}\times\nonumber\\
&&(\frac{a''r^2}{2}+\frac{a'^2r^2}{2}-\frac{a'b'r^2}{4}+a'r)f_{1R}+\frac{e^{-2b}}{r^2}\times\nonumber\\
&&\{(1-e^b)(a'^2+2a''-a'b')-2a'b')\}f_{2G}-\frac{f}{2},
\end{eqnarray}
\begin{eqnarray}\label{6}
p_r+\frac{Q^2}{8\pi r^4}e^{-(a+b)}&=&e^{-b}\big(\frac{a'}{2}+b'+\frac{2}{r}\big)f'_{1R}+e^{-2b}\{(2a''b'+a'^2b'-a'b'^2+\frac{4a'b'}{r}+\nonumber\\
&&\frac{8a'}{r^2}-\frac{4b'^2}{r}-\frac{4a'e^b}{r^2}-\frac{4b'e^b}{r^2}-\frac{8e^b}{r^3}+\frac{2a'}{r^2}+\frac{4b'}{r^2}+\frac{8}{r^3})-r\times\nonumber\\
&&(2a''b'+a'^2b'-a'b'^2-\frac{4b'^2}{r})\} f'_{2G}-\frac{e^{-b}}{r^2}(\frac{a''r^2}{2}+\frac{a'^2r^2}{4}-\frac{a'b'r^2}{4}-\nonumber\\
&&b'r)f_{1R}-\frac{e^{-2b}}{r^2}\{(1-e^b)(a'^2+2a''-a'b')- 2a'b'\}f_{2G}+\frac{f}{2},
\end{eqnarray}
\begin{eqnarray}\label{7}
p_t-\frac{Q^2}{8\pi r^4}e^{-(a+b)}&=&e^{-b} f''_{1R}+e^{-2b}(2a''+a'^2-a'b'+\frac{2a'}{r}-\frac{2b'}{r})f''_{2G}+e^{-b}(\frac{a'}{2}+\nonumber\\
&&\frac{b'}{2}+\frac{1}{r})f'_{1R}-\frac{1}{2r^3}\{e^{-b}(2b'^2r^2+4b'r-4a'^2r^2-16a'r-24-\nonumber\\
&&2a'a''r^3+a'^2b'r^3-a'^3r^3)+32-8e^b\}f'_{2G}-\frac{e^{-b}}{r^2}(\frac{a'r}{2}-\frac{b'r}{2}-\nonumber\\
&&e^b+1)f_{1R}-\frac{e^{-2b}}{r^2}\{(1-e^b)(a'^2+2a''-a'b')-2a'b'\}f_{2G}+\nonumber\\
&&\frac{f}{2}.
\end{eqnarray}
The system of three equations (\ref{5})-(\ref{7}) consists of unknown functions $\rho, p_r, p_t,$ $a, b, f$ and $E^2=\frac{Q^2}{r^2}$. This system of equations is highly non-linear and very much complicated  because of presence of bivariate function $f(R,G)$. One could find the exact solution of the system (\ref{5})-(\ref{7}) by specifying an equation of the state parameter, selecting the gravitational potentials $a$ and $b$ on the basis of physical grounds or by recommending the behavior of $\Delta$, the anisotropy parameter. For the present analysis, we choose $f(R,G)=f_1(R)+f_2(G)$. Furthermore, we consider Starobinsky like model $f_1(R)=R+\lambda R^2$, where $\lambda$ is an arbitrary constant and assume $f_2(G)=G^2$. Also $f_{1R}=\frac{df_1}{dR}$, $f_{2G}=\frac{df_2}{dG}$ and prime denotes the derivatives with respect to radial coordinate.

For the electrically charged fluid sphere, the mass function denoted by $M(r)$ is defined as
\begin{equation}
M(r)=\frac{r}{2}(1-e^{-b(r)}+\frac{Q^2}{r^2}).
\end{equation}
Here, the anisotropy measure is defined as  $\Delta=p_t-p_r$, from equations (\ref{6}) and (\ref{7}), we obtain
\begin{eqnarray}\label{8}
\Delta&=&e^{-b} f''_{1R}+e^{-2b}(2a''+a'^2-a'b'+\frac{2a'}{r}-\frac{2b'}{r})f''_{2G}-e^{-b}(\frac{b'}{2}+\frac{1}{r})f'_{1R}\nonumber\\
&&-[\frac{1}{2r^3}\{e^{-b}(2b'^2r^2+4b'r-4a'^2r^2-16a'r-24-2a'a''r^3+a'^2b'r^3-a'^3r^3)\nonumber\\
&&+32-8e^b\}+e^{-2b}\{(2a''b'+a'^2b'-a'b'^2+\frac{4a'b'}{r}+\frac{8a'}{r^2}-\frac{4b'^2}{r}-\frac{4a'e^b}{r^2}-\nonumber\\
&&\frac{4b'e^b}{r^2}-\frac{8e^b}{r^3}+\frac{2a'}{r^2}+\frac{4b'}{r^2}+\frac{8}{r^3})-r(2a''b'+a'^2b'-a'b'^2-\frac{4b'^2}{r})\}]f'_{2G}+\nonumber\\
&&\frac{e^{-b}}{r^2}(\frac{a''r^2}{2}+\frac{a'^2r^2}{4}-\frac{a'b'r^2}{4}-\frac{b'r}{2}-\frac{a'r}{2}+e^b-1)f_{1R}+\frac{Q^2}{4\pi r^4}e^{-(a+b)}.
\end{eqnarray}
The behavior of anisotropy factor has been expressed graphically in Fig. $1$. It is important to mention here that in all graphs dashed, dotted and thick curves are corresponding to $m=-10, m=4$ and $m=100000$ respectively. We note here that $\Delta=0$ represents that the inside pressure is isotropic in the the matter distribution. On the other hand, $\Delta >0 (p_t>p_r)$ and $\Delta <0(p_t<p_r)$ indicates the anisotropic force is directed outwards and inwards respectively. It is further noticed that for increasing values of $r$, $\Delta >0$ for our proposed model. This is an implication of the fact that the anisotropic force allows the structure of huge configurations. Moreover, it becomes zero at star's center.
\begin{figure}\center
\begin{tabular}{cccc}
\epsfig{file=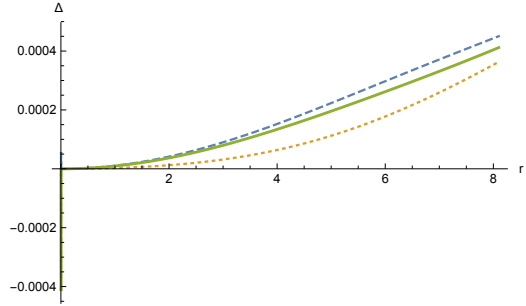,width=0.5\linewidth}
\end{tabular}
\caption{Variation of anisotropic factor $\Delta$ with radius $r$.}\center
\end{figure}

\section{The embedding class condition and generalized model for charged compact star}

Consider a $5$-dimensional flat space time
 \begin{equation}\label{8a}
 ds^2=-(dx^1)^2-(dx^2)^2-(dx^3)^2-(dx^4)^2+(dx^5)^2,
 \end{equation}
 where we assume the coordinates $x^1$, $x^2$, $x^3$, $x^4$ and $x^5$ have the following forms:
 \begin{eqnarray}\nonumber
 &&x^1=rsin\theta cos\phi,~~ x^2=rsin\theta sin\phi,~~ x^3=rcos\theta, \nonumber\\
 &&x^4=\sqrt{K}e^{\frac{a}{2}}cosh\frac{t}{\sqrt{K}},~~ x^5= \sqrt{K}e^{\frac{a}{2}}sinh\frac{t}{\sqrt{K}}.\nonumber
 \end{eqnarray}
$K$ represents a positive constant. The above defined components can be written in the differential forms as
\begin{equation}\label{9}
 dx^1=dr sin\theta cos\phi+r cos\theta cos\phi d\theta-r sin\theta sin\phi d\phi,
 \end{equation}
\begin{eqnarray}
 &&dx^2=dr sin\theta sin\phi+r cos\theta sin\phi d\theta+r sin\theta cos\phi d\phi,\\
 &&dx^3=dr cos\theta-r sin\theta d\theta,\\
 &&dx^4=\sqrt{K}e^{\frac{a}{2}}\frac{a'}{2}cosh\frac{t}{\sqrt{K}}dr+e^{\frac{a}{2}}sinh\frac{t}{\sqrt{K}}dt,\\
 &&dx^5=\sqrt{K}e^{\frac{a}{2}}\frac{a'}{2}sinh\frac{t}{\sqrt{K}}dr+e^{\frac{a}{2}}cosh\frac{t}{\sqrt{K}}dt,\label{10}
\end{eqnarray}
where prime stands for the derivative with respect to $r$.
Putting all expressions (\ref{9})-(\ref{10}) in metric (\ref{8a}), we get:
\begin{equation}\label{11}
 ds^2=-(1+\frac{Ke^a}{4}a'^2)dr^2-r^2(d\theta^2+sin^2\theta d\phi^2)+e^{a(r)}dt^2.
 \end{equation}
 From the comparison of metric (\ref{1}) and metric (\ref{11}), we obtain:
 \begin{equation}\label{12}
 e^b=(1+\frac{Ke^a}{4}a'^2).
 \end{equation}
 The equation (\ref{12}) gives the embedding class one condition.
In recent years, researchers have focused their much attention in modeling of compact stellar like neutron and strange stars or pulsars. This is because of the fact that huge observational data is available now a days against which the merits and demerits of different theoretical models may be examined.

Now, we discuss a general model by assuming a  monotone increasing generic function $a(r)$. The invariance of Ricci tensor needs that all three quantities radial pressure, tangential pressure and energy density must be finite at the origin. Furthermore, both charge function $Q(r)$ and mass function $M(r)$ achieve minimum values at $r=0$, center of the structure, and get maximum values at the surface. Mathematically speaking, $Q(0)=0,$ $Q'(0)>0$ and $M(0)=0, M'(0)>0$. Keeping all these observations, we assume the following form of the generic function $a(r)$ \cite{15a}.
\begin{equation}\label{13}
a(r)=m~~ln(1+Ar^2)+lnB.
\end{equation}
Where $A$ and $B$ are constants and we further consider two cases.

Case $(i)$ $m<0$ and $A<0$, Case $(ii)$ $m>0$ and $A>0$.

For both these above mentioned cases, we note that $a(0)=lnB,  a'=\frac{2mAr}{(1+Ar^2)}$ and $a''=\frac{2m(1-Ar^2)}{(1+Ar^2)^2}$. It implies that $a(0)>0, a'(0)=0, a''(0)=2mAr>0$ and $a(r)\neq0$ with $r\neq0$. So it provides that $a(r)$ is an increasing function having regular minimum value at $r=0$. Putting this value of $a$ into equation (\ref{12}), we obtain
\begin{equation}\label{14}
b=ln[1+CAr^2(1+Ar^2)^{(m-2)}],
\end{equation}
where $C=m^2ABK$.

The arrangement of the metric potential function (\ref{14}) has been used by many researchers to model compact objects in the context of Karmarkar condition \cite{15c}. The rule of parameter $m$ is very important in the analysis of stability and configuration of the compact objects. There is a noteworthy link between the range of $m$ and physically feasible models. It is clear from equation (\ref{13}) that space time reduced flat for vanishing $m$. We will discuss solutions for $m<0$ and $m>0$. This approach permits one to examine the impact of $m$  on several thermodynamical characteristics of the model. We will describe compact star $Her X-1$ for $m\leq-7$ and $m\geq4$ because outside of this range solution is not well behaved \cite{15a}.

The charge function $Q(r)$ is defined here as:
\begin{equation}\label{14a}
Q(r)=E_0A^2r^6(1+Ar^2)^m,
\end{equation}
with $E_0$ as positive constant. For physical suitable model having charged anisotropic fluid, both metric potentials should be non zero positive in the range $0\leq r \leq R$ or equivalently, $0\leq r/R \leq 1$. Fig. $2$ shows that this condition is true here. From Fig. $3$ and Fig. $4$, we observe $M(0)=Q(0)=0$ at $r=0$. Also the derivative of both functions are positive when $0\leq\frac{r}{R}\leq1$, which indicates the monotonically increasing behavior away from $r=0$, the center, and minimum value at $r=0$.
\begin{figure}\center
\begin{tabular}{cccc}
\epsfig{file=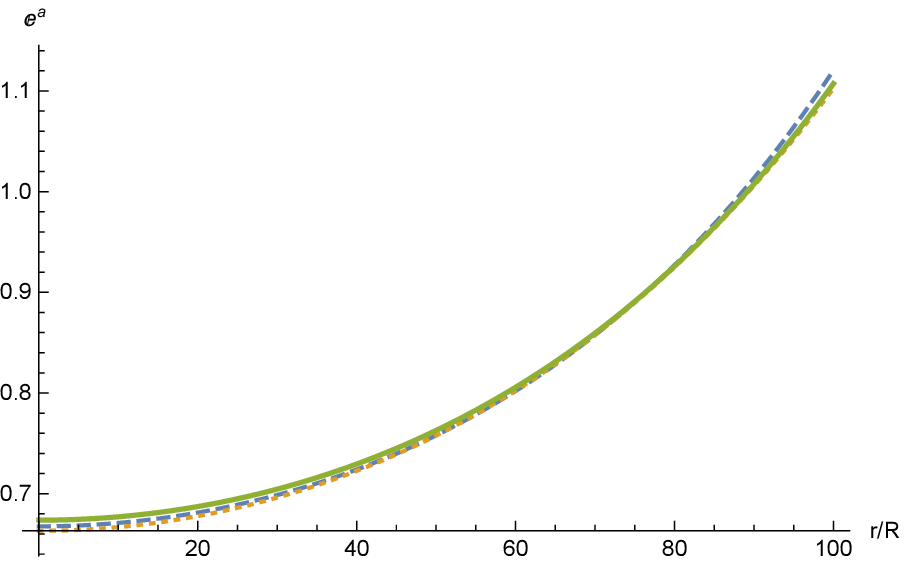,width=0.35\linewidth} &
\epsfig{file=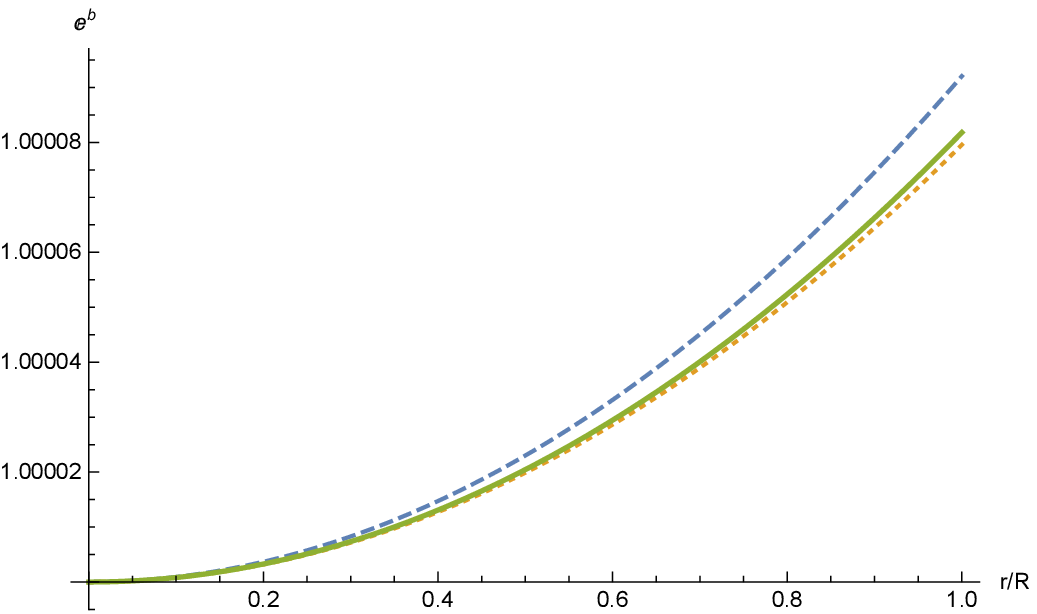,width=0.35\linewidth} \\
\end{tabular}
\caption{Variation of gravitational potentials $e^a$ and $e^b$ with fractional radius $r/R$ for  $m=-10, m=4, m=100000$ for Her $X-1$. }\center
\end{figure}
\begin{figure}\center
\begin{tabular}{cccc}
\epsfig{file=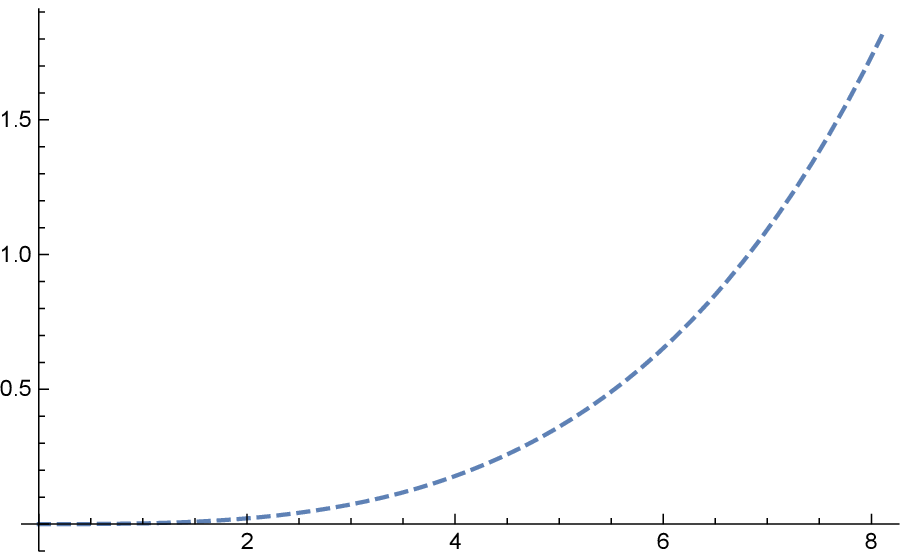,width=0.35\linewidth} &
\epsfig{file=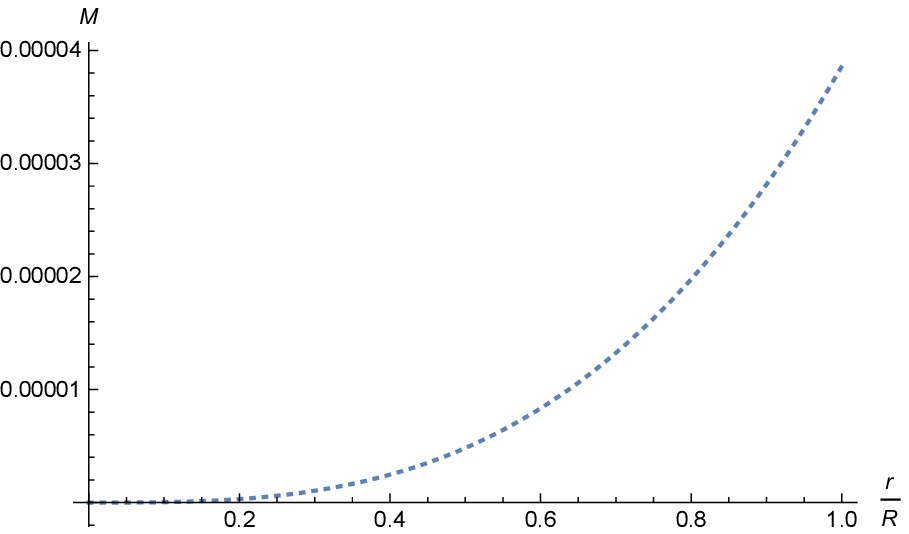,width=0.35\linewidth} &
\epsfig{file=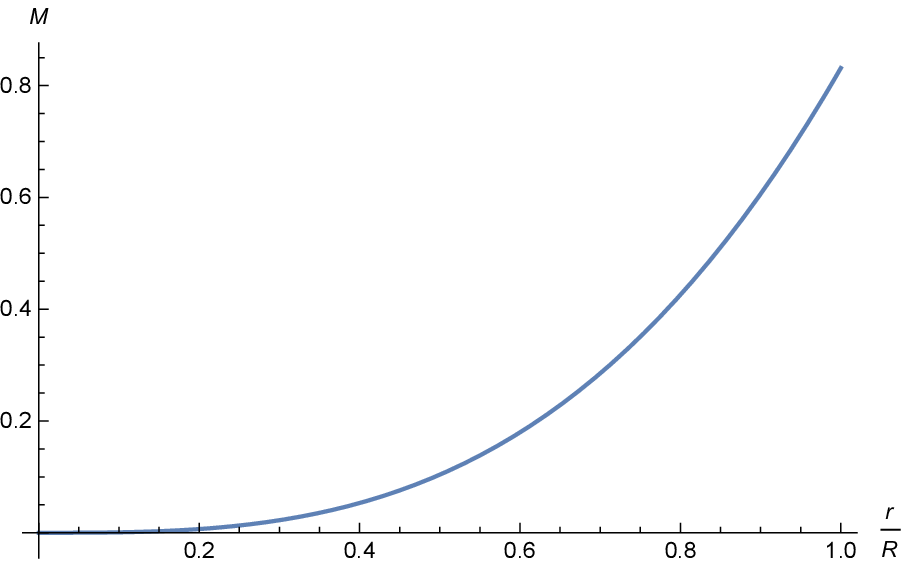,width=0.35\linewidth} \\
\end{tabular}
\caption{Variation of mass function $M(r)$ with fractional radius $r/R$ for  $m=-10$ (left curve), $m=4$ (middle curve) and $m=100000$ (right curve) for Her $X-1$.}\center
\end{figure}
\begin{figure}\center
\begin{tabular}{cccc}
\epsfig{file=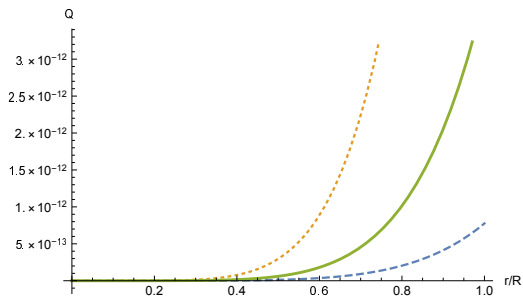,width=0.5\linewidth}
\end{tabular}
\caption{Variation of Electric charge $Q(r)$ with fractional radius $r/R$ for Her $X-1$.}\center
\end{figure}
\begin{figure}\center
\begin{tabular}{cccc}
\epsfig{file=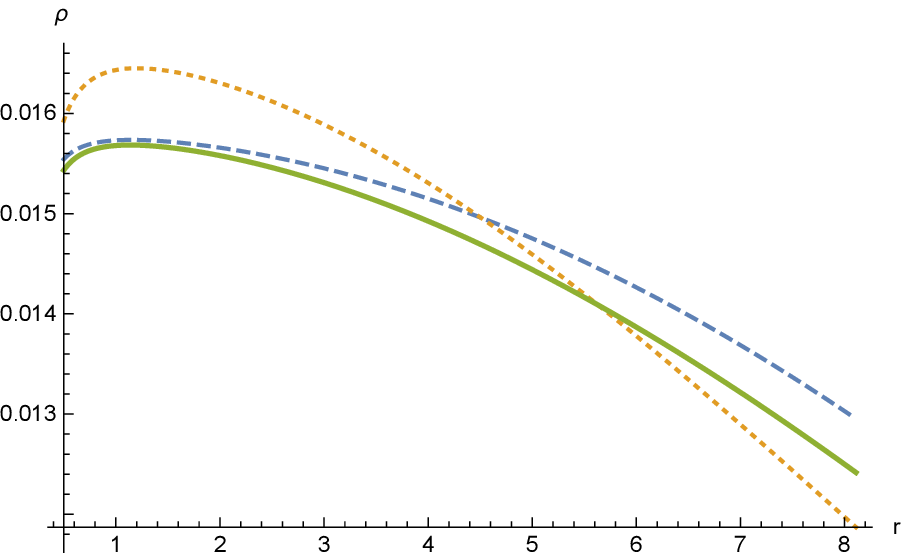,width=0.35\linewidth} &
\epsfig{file=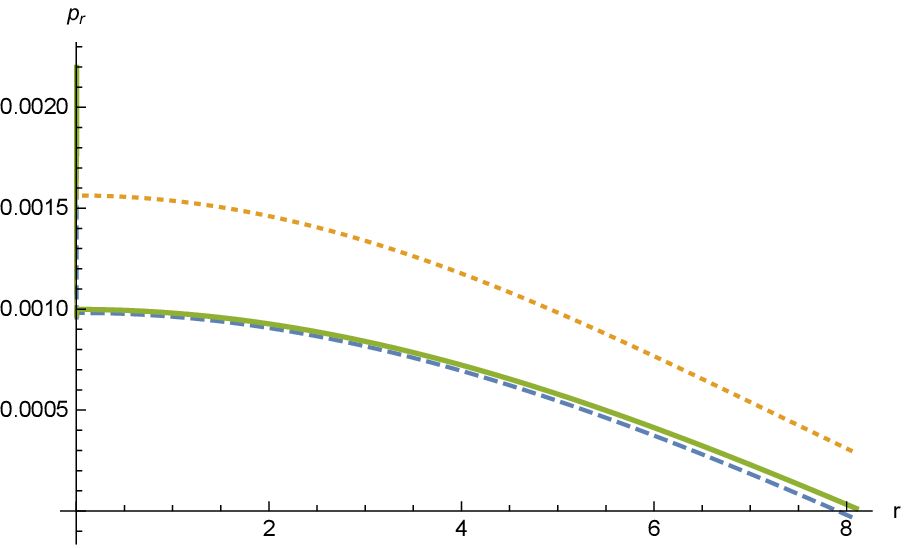,width=0.35\linewidth} &
\epsfig{file=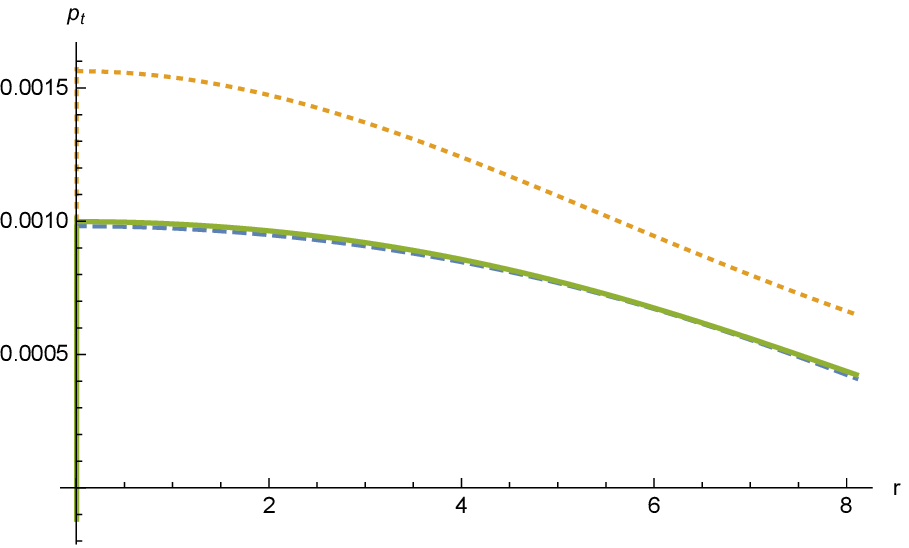,width=0.35\linewidth} \\
\end{tabular}
\caption{Behavior of energy density $\rho$, radial pressure $p_r$ and tangential pressure $p_t$ with radial coordinate $r$ for Her $X-1$.}\center
\end{figure}

\section{Physical analysis of the solution}
The energy density, radial and tangential pressure should be positively finite at the center of the configuration i.e $\rho>o, p_r>0$ and $p_t>0$ when $r=0$. Also energy density and pressure both should attain maximum value at the center and continuously decrease afterwards. The behavior of $\rho, p_r$ and $p_t$ has been shown in Fig. $5$ and it confirms that above mentioned requirements are fulfilled in current analysis. In addition to that we notice that
$\frac{d\rho}{dr}=0,\frac{d^2\rho}{dr^2}<0$ and $\frac{dp_r}{dr}=0,\frac{d^2p_r}{dr^2}<0$ for $r=0$. It implies that $\rho$ and $p_r$ have maximum value at the center.

\subsection{Junction conditions}
For generating physically possible confined object model, we must make sure that the interior space time $\textit{M}^-$  should smoothly match with the exterior space time $\textit{M}^+$. Reissner Nordstrom solution is considered for  $\textit{M}^+$, because exterior space time is empty.\\
The condition $e^{a(R)}=e^{-b(R)}$ yields the constant $B$ as
\begin{equation}
B=\frac{1}{(1+AR^2)^m[1+CAR^2(1+AR^2)^{m-2}]}.
\end{equation}
Star's surface density can be used to determine the constant $A$.

The radial pressure disappears at the boundary of the object. i.e. $p_r=0$ at $r=R$. The constant $C$ is obtained by using the condition $p_r=0$.\\
Using the estimated values of mass $M=0.88M_{\odot}$ and radius $R=8.10$ $km$ for the compact star $Her X-1$, the constants $AR^2, A, C, B$ and $E_0$ are shown in the following table \cite{16}
\begin{center}
\begin{tabular}{ |c|c|c|c|c|c|c| }
 \hline
 m  &      $AR^2$           & $A$  & $C$& $B$     & $K$   & $E_0$ \\
 \hline
 \textbf{$-10$ } & $-0.0217$  & $-0.3311\times10^{-3}$ &$-15.7670$ & $0.6676$ & $7.1328\times10^2$ & $0.0200\times10^2$  \\
 \hline
 \textbf{$4$}  & $0.0583$  & $0.889\times10^{-3}$  &$6.79354$& $0.6632$ & $7.2012\times10^2$ & $0.48$   \\
 \hline
 \textbf{$100000$}& $0.00000214$& $0.3260\times10^{-7}$ &$1.6463\times10^5$& $0.6737$ &$7.4960\times10^2$ & $0.1027\times10^{10}$  \\
 \hline
\end{tabular}
\end{center}
\subsection{Energy conditions}
In study and discussion of some significant cosmological issues, energy conditions are considered very useful. Many important consequences have been described using these energy conditions \cite{16a}-\cite{16d}.
Inside the charged anisotropic fluid sphere, the following energy bounds must be satisfied simultaneously
\begin{equation}
NEC:\rho+\frac{Q^2}{8\pi r^4}\geq0,
\end{equation}
\begin{equation}
WEC: \rho+p_r\geq0,~~~~~~\rho+p_t+\frac{Q^2}{4\pi r^4}\geq0,
\end{equation}
\begin{equation}
SEC: \rho+p_r+2p_t+\frac{Q^2}{4\pi r^4}\geq0,
\end{equation}
Behavior of above mentioned energy conditions is shown in Fig. $6$ and it is clear from the graphs that these conditions are well satisfied in the range $0\leq r \leq R.$
\begin{figure}\center
\begin{tabular}{cccc}
\epsfig{file=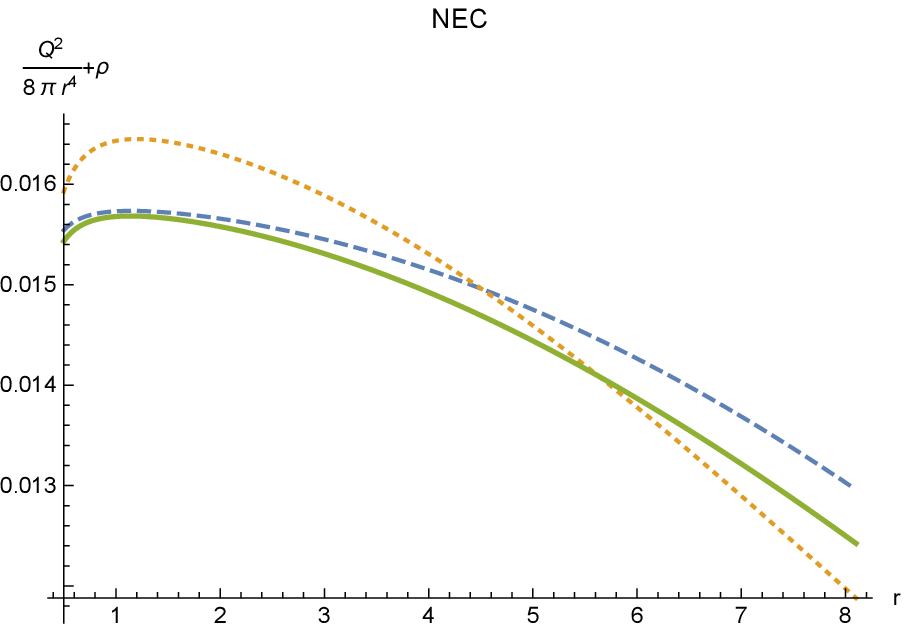,width=0.35\linewidth} &
\epsfig{file=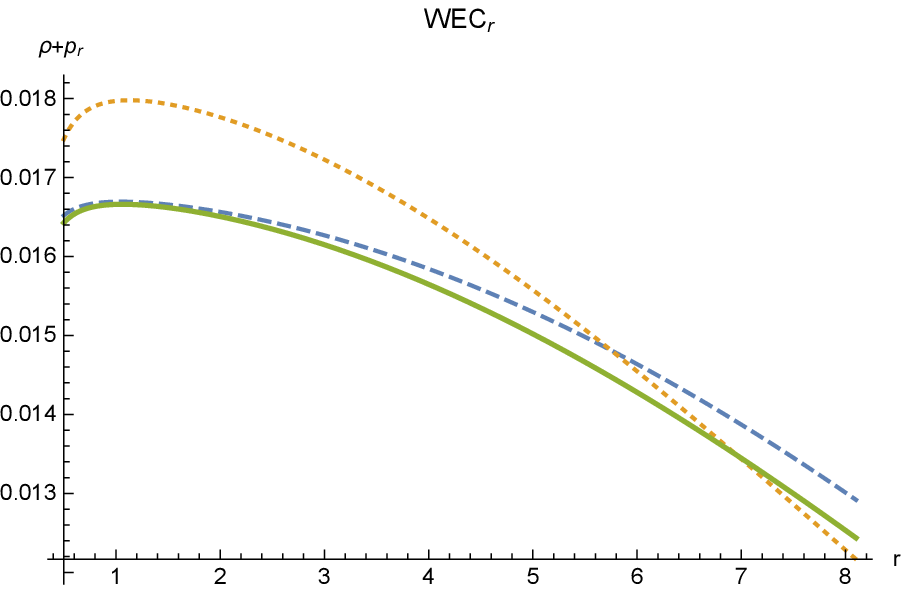,width=0.35\linewidth} &\\
\epsfig{file=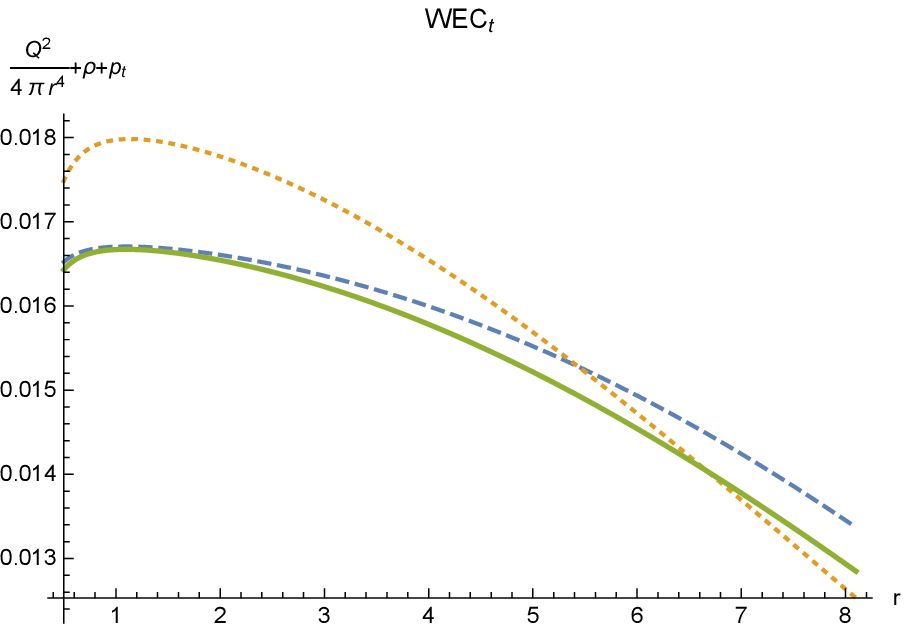,width=0.35\linewidth} &
\epsfig{file=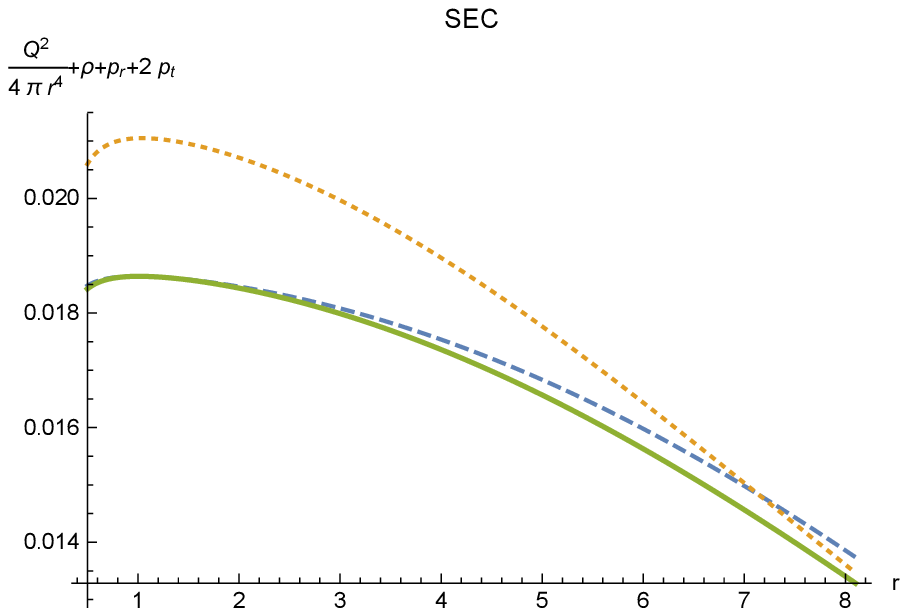,width=0.35\linewidth} &\\
\end{tabular}
\caption{Behavior of energy Conditions for compact star $Her X-1$ with radial coordinate $r$}\center
\end{figure}

\subsection{Equilibrium condition using Tolman Oppenheimer Volkoff (TOV) equation}

The TOV equation including charge states that the connection between all four forces, anisotropic force $(F_a)$, gravitational force $(F_g)$, hydrostatic force $(F_h)$ and electric force $(F_e)$ in the form of the following equation \cite{17}, \cite{18}
\begin{equation}
\frac{dp_r}{dr}=\frac{2\Delta}{r}+\frac{Q}{4\pi r^4}\frac{dQ}{dr}-Br(\rho+p_r),
\end{equation}
where
\begin{equation}
F_a=\frac{2\Delta}{r}, F_g=-Br(\rho+p_r), F_h=-\frac{dp_r}{dr}, F_e=\frac{Q}{4\pi r^4}\frac{dQ}{dr}.
\end{equation}
\begin{figure}\center
\begin{tabular}{cccc}
\epsfig{file=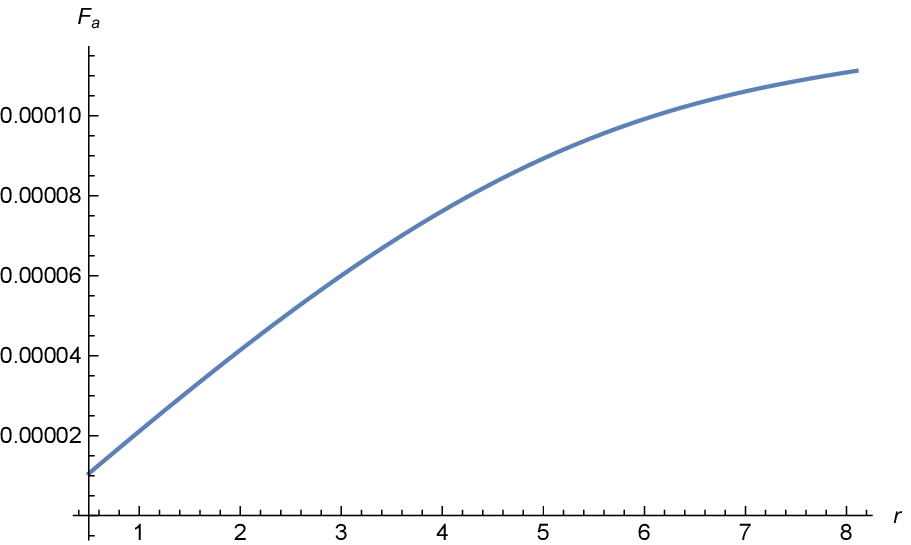,width=0.25\linewidth} &
\epsfig{file=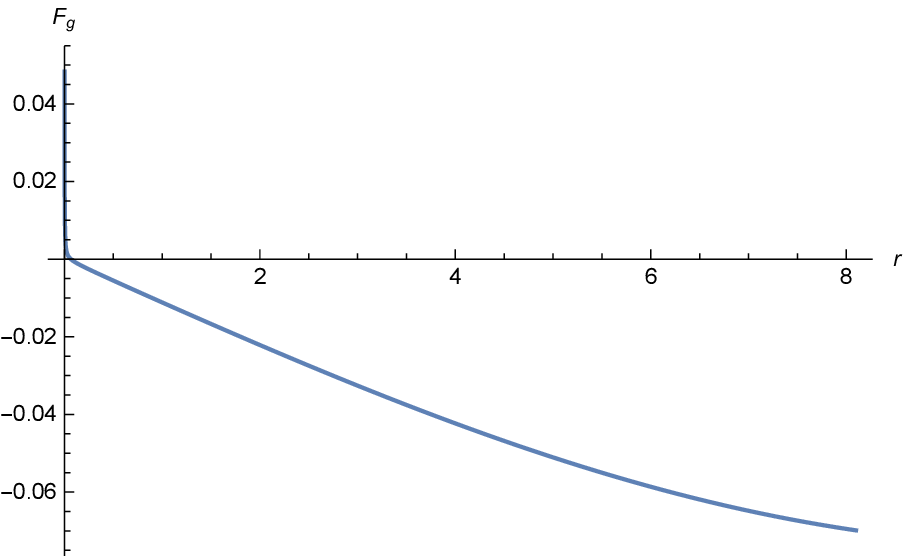,width=0.25\linewidth} &
\epsfig{file=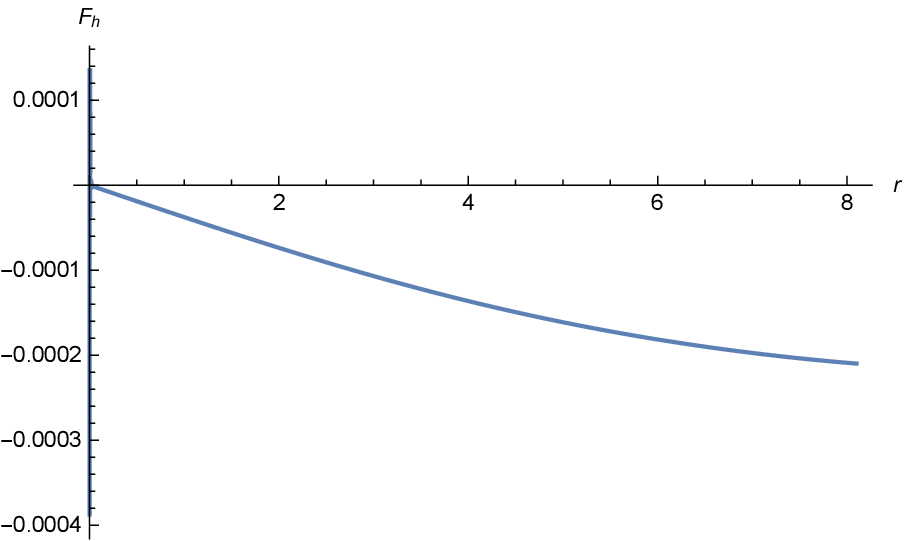,width=0.25\linewidth} &
\epsfig{file=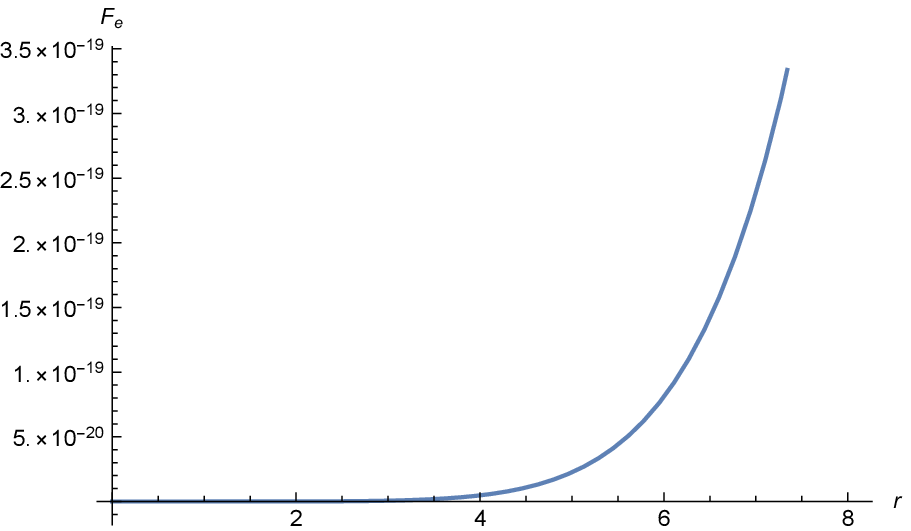,width=0.25\linewidth} \\
\end{tabular}
\caption{Behavior all four forces $F_a, F_g, F_h, F_e$ versus $r$ for $m=-10$ .}\center
\end{figure}
\begin{figure}\center
\begin{tabular}{cccc}
\epsfig{file=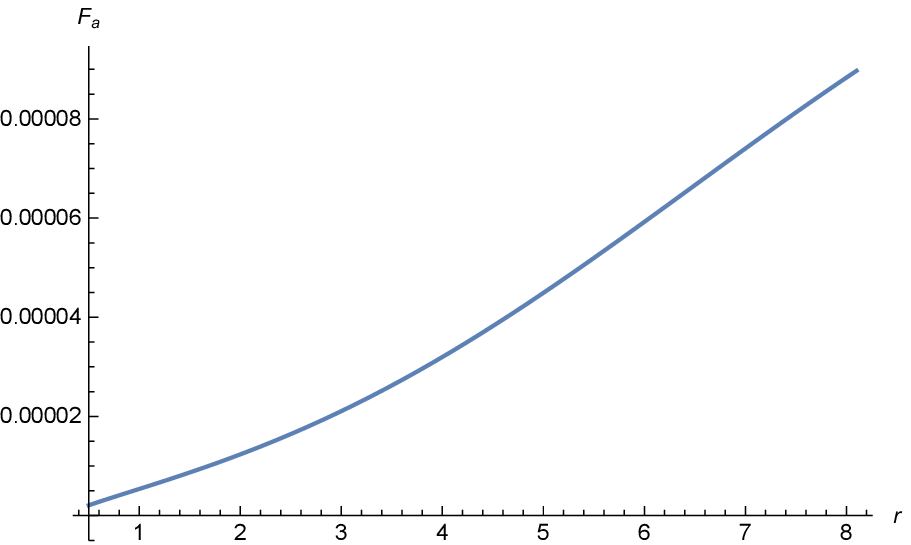,width=0.25\linewidth} &
\epsfig{file=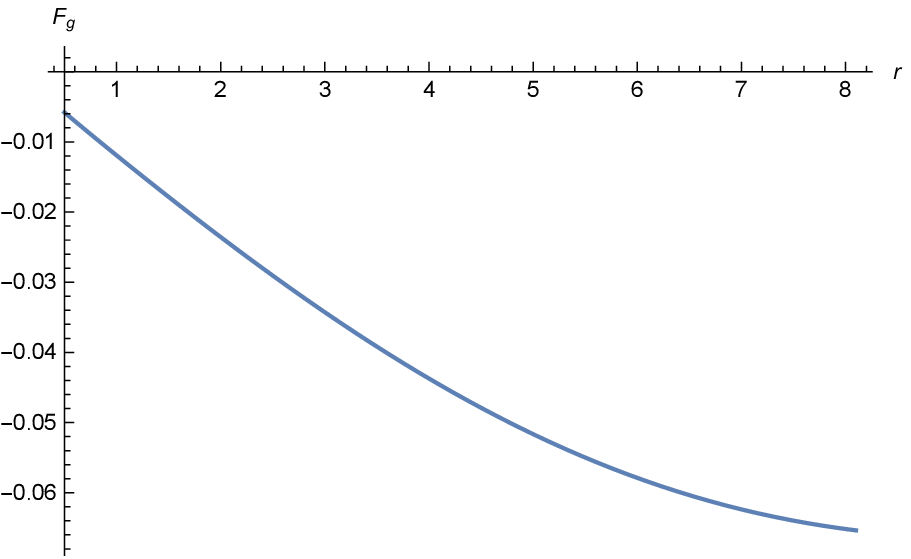,width=0.25\linewidth} &
\epsfig{file=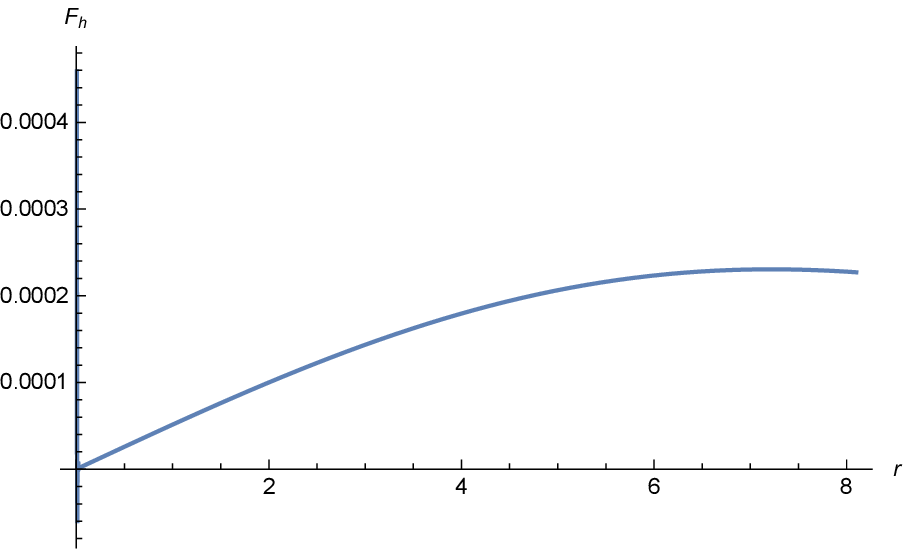,width=0.25\linewidth} &
\epsfig{file=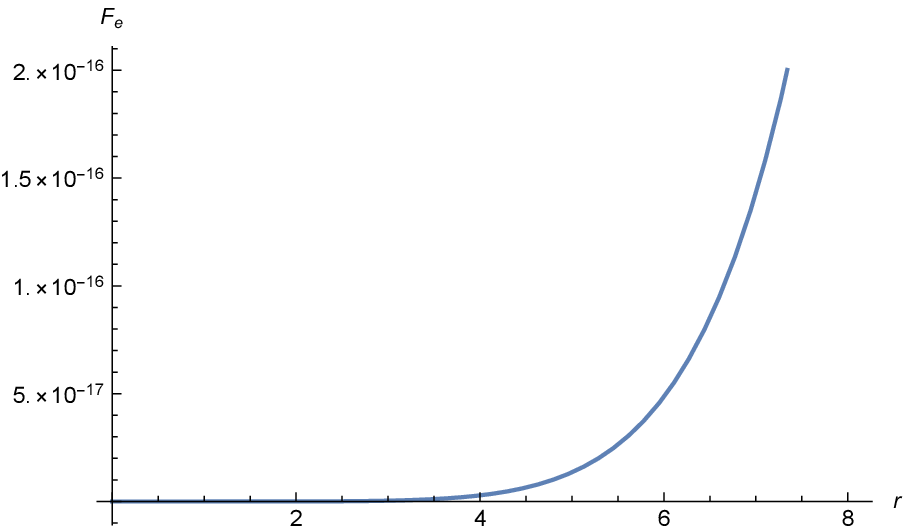,width=0.25\linewidth} \\
\end{tabular}
\caption{Behavior all four forces $F_a, F_g, F_h, F_e$ versus $r$ for $m=4$ .}\center
\end{figure}

\begin{figure}\center
\begin{tabular}{cccc}
\epsfig{file=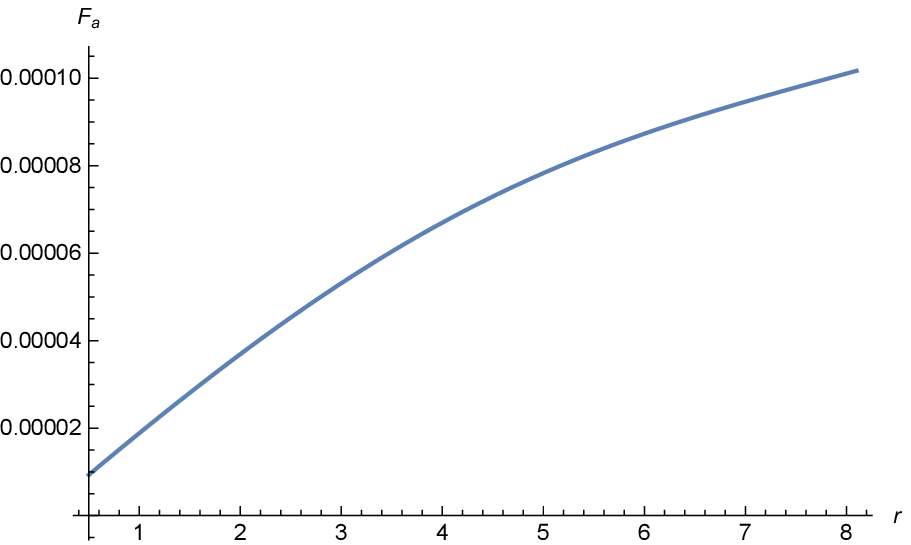,width=0.25\linewidth} &
\epsfig{file=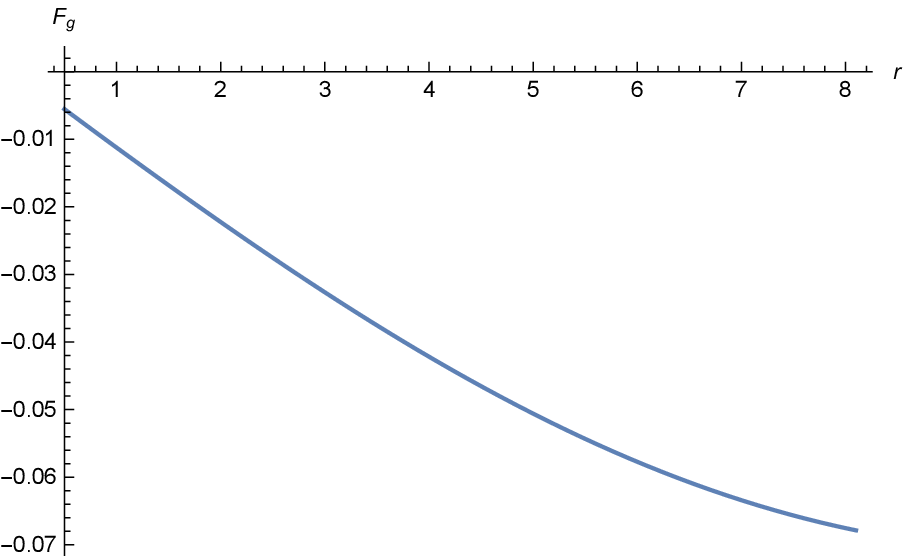,width=0.25\linewidth} &
\epsfig{file=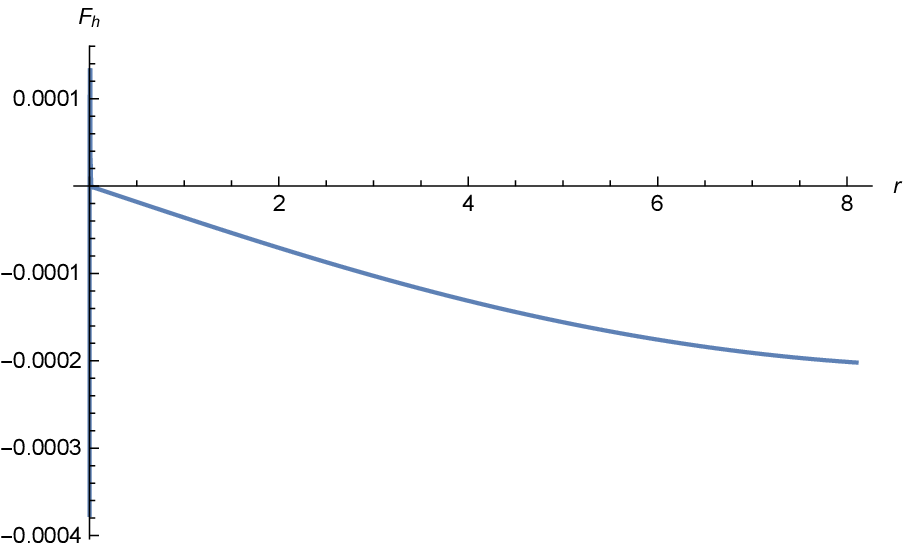,width=0.25\linewidth} &
\epsfig{file=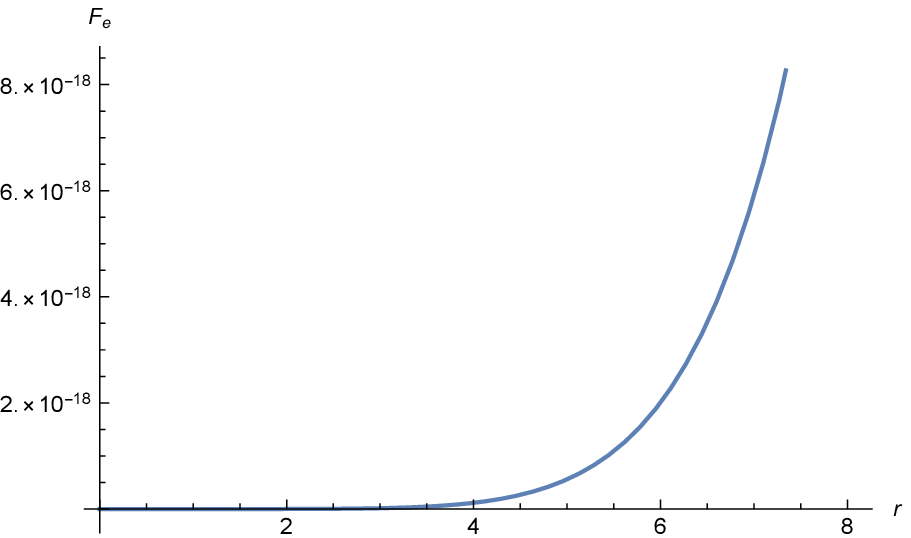,width=0.25\linewidth} \\
\end{tabular}
\caption{Behavior all four forces $F_a, F_g, F_h, F_e$ versus $r$ for $m=100000$.}\center
\end{figure}

Observations from different panels in Figs. $(7, 8, 9)$ indicate that anisotropic force dominates the electric force and behavior of gravitational force remains almost same for different values of $m$. Hydrostatic force increases for small values of $m$ and decreases for the large values of $m$.

\subsection{Stability Analysis}

For the stability analysis of our model, we observe the variation of both radial sound speed $(v_{sr})$ and transverse sound speed $(v_{st})$. It is clear from Fig. $(10)$ that both inequalities $0\leq v^2_{sr}\leq 1$ and $0\leq v^2_{st}\leq 1$ are satisfied inside the star which is an indication of causality preservation within the star. It can also be noticed that both sound speeds show monotonically decreasing trend from high density area (star's center) towards low density region (star's surface).

Furthermore, we use Herrera's cracking idea to determine stability of charged distribution with anisotropic matter \cite{19}. He perturbed the energy density and anisotropy and studied the consequences of perturbations inside fluid components. He was of the view that various portions of the star react in a different way to several amounts of anisotropy.  This behavior may be an indication of cracking within the core. Abreu et al. \cite{109} studied the cracking within the static sphere with another approach. They utilized the radial sound speeds and tangential sound speeds for indication of unstable regions. It requires that $\mid v^2_{sr}-v^2_{st}\mid \leq 1$ for possibly stable region. This is shown in Fig. $(11)$ that this inequality is true in our case. So, our model is stable.
\begin{figure}\center
\begin{tabular}{cccc}
\epsfig{file=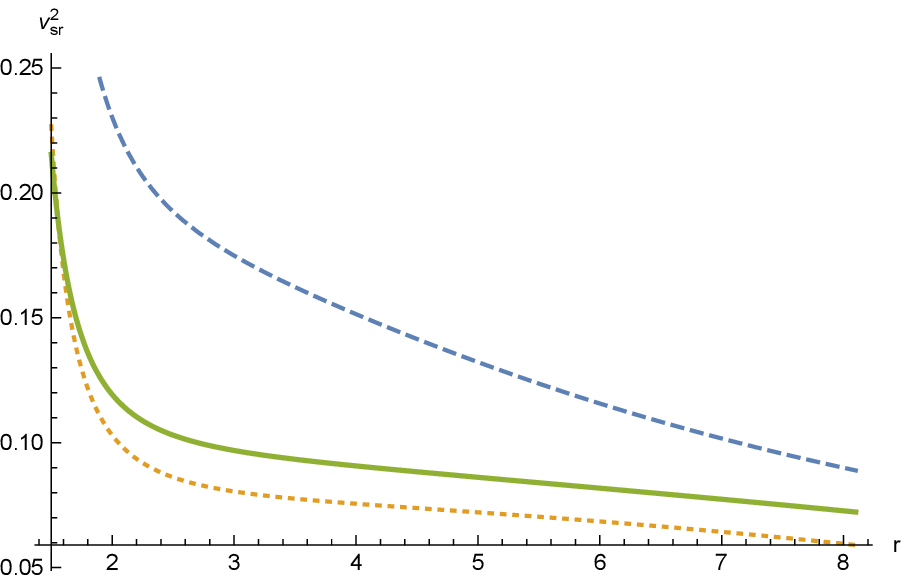,width=0.35\linewidth} &
\epsfig{file=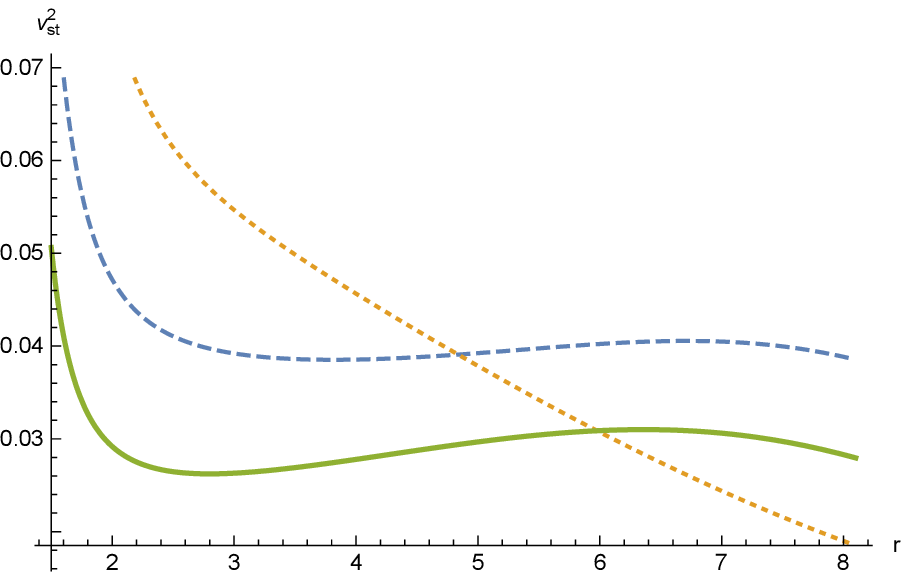,width=0.35\linewidth} \\
\end{tabular}
\caption{Variation of radial sound speed $v^2_{sr}$ and tangential sound speed$ v^2_{st}$ versus $r$}\center
\end{figure}
\begin{figure}\center
\begin{tabular}{cccc}
\epsfig{file=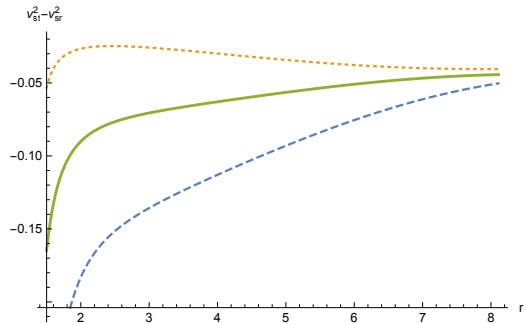,width=0.50\linewidth}
\end{tabular}
\caption{Variation of the expression $v^2_{sr}-v^2_{st}$ versus $r$}\center
\end{figure}

\section{Concluding remarks}

We have focused on spherically symmetric charged anisotropic configuration having internal geometry of embedding class one with a proposed model of $f(R,G)$ gravity. The following important physical features have been noticed.\\\\
\begin{itemize}
\item Fig. $1$ shows the behavior of gravitational potentials versus fractional radius $r/R$. The continuous gravitational potentials display the smoothly increasing trend from the center to the surface of the star. An increase or decrease in $m$ has no Significant effect on the nature of these gravitational potentials.
\item The mass and the charge functions are zero at the center and increase towards star's surface. Increasing the value $m$ also creates the monotonic increase in mass and charge. The evolution of energy density, radial pressure and tangential pressure show an ever decreasing behavior. The value of $m$ has very little effect in the behavior of all three quantities as we have examined these for $m=-10, 4, 100000$. The density profile attains maximum value at the center of the star and it remains almost same for even very large value of $m$ everywhere inside the star.
\item The anisotropy measure vanishes at $r=0$ and becomes monotonically increasing outwards near the surface of the star.
\item All the energy conditions are found to be true in the interior of the star.
\item Different forces working within the interior are graphically presented in Figs. $(7, 8, 9)$. Three forces, anisotropic force, gravitational force and electric force show the same behavior for different values of $m$ while forth force namely, hydrostatic force, shows opposite trend for $m=4$ as compare to the other two values $m=-10$ and $m=100000$.
\item Stability of our proposed model is declared by finding the relative difference of squared radial sound speed and tangential sound speed as it was found to be less than unity. i.e. $\mid v^2_{sr}-v^2_{st}\mid \leq 1$.\\\\
\end{itemize}
For the present analysis we have also chosen a very large value of $m$, i.e. $m=100000$. The results indicate that the physical conditions behave usually as in the case of small values of $m$. Thus, it is conjectured that large values of $m$ will have no greater effect and the results will not change.
It is important to mention here that the results of this paper confirms the findings in \cite{14} for $Q=0$.

\end{document}